\documentclass[conference]{IEEEtran}
\IEEEoverridecommandlockouts

\usepackage{multirow}
\usepackage{cite}
\usepackage{amsmath,amssymb,amsfonts}
\usepackage{algorithm}
\usepackage{algorithmic}
\usepackage{graphicx}
\usepackage{textcomp}
\usepackage{xcolor}
\usepackage{booktabs}
\usepackage{authblk}
\usepackage{amssymb}
\usepackage{pifont}
\usepackage{url}
\newcommand{\crossmark}{\ding{61}}%

\def\BibTeX{{\rm B\kern-.05em{\sc i\kern-.025em b}\kern-.08em
    T\kern-.1667em\lower.7ex\hbox{E}\kern-.125emX}}

\begin{document}

\title{Leveraging Frequency Domain Learning in 3D Vessel Segmentation
\thanks{Corresponding author: Chengwei Pan (pancw@buaa.edu.cn). This work was supported by the National Key R\&D Program of China (2022ZD0116401).}
}

\author[1]{\textbf{Xinyuan Wang}}
\author[1,3]{\textbf{Chengwei Pan}}
\author[4]{\textbf{Hongming Dai}}
\author[5,6]{\textbf{Gangming Zhao}}
\author[7,8]{\textbf{Jinpeng Li}}
\author[2,3]{\\ \textbf{Xiao Zhang}}
\author[5,6]{\textbf{Yizhou Yu}}

\affil[1]{Institute of Artificial Intelligence, Beihang University, Beijing, China}
\affil[2]{School of Mathematical Sciences, Beihang University, Beijing, China}
\affil[3]{Zhongguancun Laboratory, Beijing, China}
\affil[4]{School Of Computing, National University of Singapore}
\affil[5]{Department of Computer Science, University of Hong Kong, Hong Kong, China}
\affil[6]{Deepwise AI Lab, Beijing, China}
\affil[7]{HwaMei Hospital, University of Chinese Academy of Sciences (UCAS), Ningbo, China}
\affil[8]{Ningbo Institute of Life and Health Industry, UCAS, Ningbo, China}

\maketitle

\begin{abstract}

Coronary microvascular disease constitutes a substantial risk to human health. Employing computer-aided analysis and diagnostic systems, medical professionals can intervene early in disease progression, with 3D vessel segmentation serving as a crucial component. Nevertheless, conventional U-Net architectures tend to yield incoherent and imprecise segmentation outcomes, particularly for small vessel structures. While models with attention mechanisms, such as Transformers and large convolutional kernels, demonstrate superior performance, their extensive computational demands during training and inference lead to increased time complexity. 
In this study, we leverage Fourier domain learning as a substitute for multi-scale convolutional kernels in 3D hierarchical segmentation models, which can reduce computational expenses while preserving global receptive fields within the network.  Furthermore, a zero-parameter frequency domain fusion method is designed to improve the skip connections in U-Net architecture. Experimental results on a public dataset and an in-house dataset indicate that our novel Fourier transformation-based network achieves remarkable dice performance (84.37\% on ASACA500 and 80.32\% on ImageCAS) in tubular vessel segmentation tasks and substantially
reduces computational requirements without compromising global receptive fields. 

\end{abstract}

\IEEEpeerreviewmaketitle


\section{Introduction}
Coronary microvascular disease is a major threat to human health. Computed Tomography Angiography (CTA) is widely used for the diagnosis and treatment planning of coronary artery disease due to its non-invasiveness and capability to provide high-resolution 3D imaging. Automatic segmentation of the coronary arteries is highly desirable to help radiologists intervene early thus improving diagnostic efficiency.

Over the years, numerous methods have been proposed for medical image segmentation. UNet\cite{ronneberger2015u} with an  encoder-decoder architecture has been widely used and given rise to a variety of variants such as 3D UNet\cite{cciccek20163d}, UNet++\cite{zhou2019unet++} and nnUNet\cite{isensee2021nnu}. However, the stacked convolutions in UNet family are hard to capture long-range dependencies between different regions, which may lead to  inaccurate segmentation due to the intricate tubular structure of the coronary arteries. Most recently, transformer models based on self-attention mechanisms have shown significant advancements\cite{chen2021transunet} within the capability of learning long-range dependencies, while the computational demands are enormous, especially in 3D segmentation. Moreover, while adept at extracting low-frequency information such as global shapes and structures, transformers may not adequately capture high-frequency elements\cite{si2022inception}. Thus it is critical to design deep neural networks that can take advantage of both low-frequency and high-frequency information simultaneously.

In this paper, we propose a method based on frequency domain learning to cover all the frequencies, and summarize contributions below:

\begin{itemize}
    \item 
    We present a 3D segmentation approach using frequency domain learning to enhance network fitting of vascular shapes by introducing global receptive field. We analyze and address aliasing from parameterized multiplication in frequency domain. Leveraging FFT efficiency significantly reduces computational load compared to attention mechanisms.
    \item 
    We propose a parameter-free skip connection strategy fusing high and low frequency domains to better integrate encoder and decoder features. Unlike a meticulously designed encoder with original skip connections, this facilitates preserving high-frequency edge features from encoder and low-frequency semantic features from decoder.
    \item 
    Extensive experiments show our approach achieves state-of-the-art 3D vessel segmentation performance on two coronary vessel datasets.

\end{itemize}

\section{Related Work} \label{sec:related}

\subsection{Vessel Segmentation}
Vessel segmentation plays an important role in medical image segmentation. Tetteh et al.\cite{tetteh2020deepvesselnet} proposed a novel convolutional neural network that can be simultaneously used for vessel segmentation, centerline extraction, and bifurcation point detection. They used three orthogonal 2D convolutional operations to replace 3D convolutional operations in order to reduce parameters and computational complexity. Wang et al.\cite{wang2020deep} introduce the deep distance transform (DDT) as a method for segmenting tubular structures in CT scans. Zeng et al.\cite{zeng2022imagecas} released a coronary segmentation dataset containing 1000 CTA images and proposed a strong baseline method using multi-scale block fusion and two-stage post-processing to capture vascular details.  Additionally, shape prior knowledge can also be introduced into networks. For example, Lee et al.\cite{lee2019tetris} introduced explicit tubular structure priors into vessel segmentation using a template deformation network. This approach is based on network registration to deform the shape template, achieving precise segmentation of input images while maintaining topological constraints. Recently, Wolterink et al.\cite{wolterink2019graph} incorporated graph convolutional networks into coronary artery segmentation tasks, treating the vertices on the surface mesh of the coronary artery lumen as graph nodes and directly optimizing the positions of these mesh vertices. Zhao et al.\cite{zhao2022graph} proposed a cross-network multi-scale feature fusion framework that utilizes the fusion of graph convolutional networks and CNN to obtain high-quality vascular segmentation results.

\subsection{Learning in Frequency Domain}
In recent years, the field of computer vision has seen a surge of interest in learning in the frequency domain, with each study building upon the successes of its predecessors. Zequn Qin's FcaNet\cite{qin2021fcanet} shifted its focus to frequency channel attention mechanisms, achieving significant performance improvements in image classification, object detection, and instance segmentation tasks relative to existing channel attention approaches. Lastly, building upon these advancements, Yongming Rao's Global Filter Network (GFNet) \cite{rao2021global} introduced a computationally efficient architecture that learns long-term spatial dependencies in the frequency domain with log-linear complexity. 

\section{Methods} \label{sec:method}

\begin{figure}[h]
  \centering
  \includegraphics[width=\linewidth]{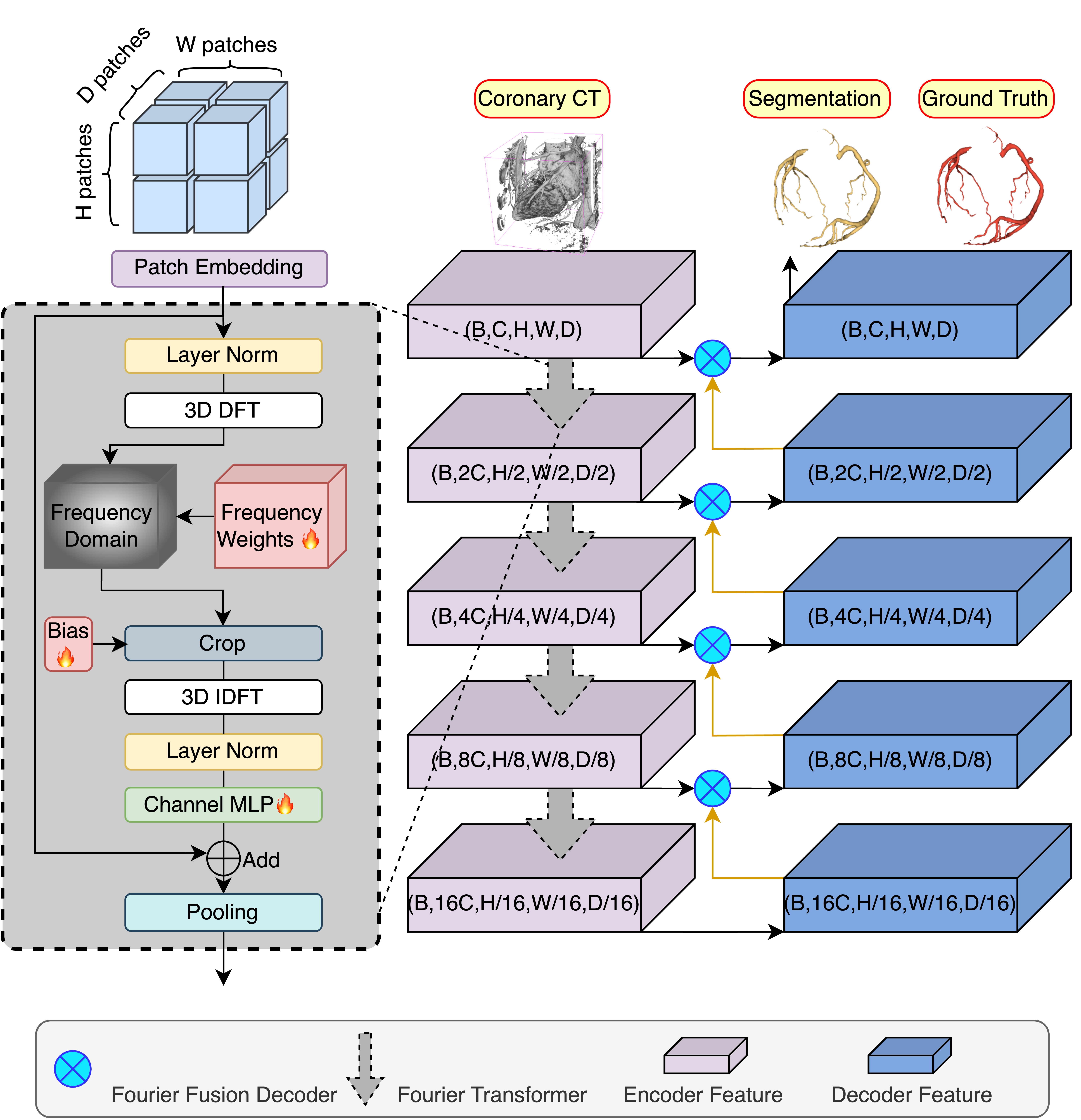}
  \caption{Hierarchical Fourier Segmentation(Fseg) Network Overview}
  \label{fig:pipeline}
\end{figure}

\subsection{Overview}

The architecture of the proposed method is presented in Fig. \ref{fig:pipeline}. Consider an input 3D image volume $X\in \mathbb{R}^{D \times H \times W}$, $D$, $H$, and $W$ represent the spatial depth, height, and width, respectively. The 3D UXNet\cite{lee20223d} is used as the backbone, which includes an encoder and a decoder. Firstly, a large-kernel convolution layer is used to extract patch-wise features as the encoder’s inputs. Then the encoder is composed of  four hierarchical stages of transformer blocks, in which the attention mechanism is replaced by a frequency-domain global weighting operation. Finally, a frequency-domain Fourier Fusion Decoder module is designed to perform the interaction between the higher-resolution features from the encoder and the lower-resolution features from the decoder.

\subsection{Discrete Fourier Transform}
Discrete Fourier Transform (DFT) plays an important role in the field of computer image processing. DFT decomposes a complex signal into single frequency components with different amplitudes, thus enabling operations such as filtering. It seems more meaningful to transform spatial domain information to the frequency domain for operations, since abstract semantic image features are generally low-frequency.

\subsubsection{3D-DFT}
Similar to one-dimensional sequences, multidimensional sequences also possess Discrete Fourier Transform (DFT). Given a $d$-dimensional sequence $x[n_1, n_2, n_3]$ with $N_d$ values in each dimension, its multidimensional DFT is given by: 

\begin{equation}
X[k_1, k_2, k_3] = \sum_{n_1 = 0}^{N_1 - 1} \sum_{n_2 = 0}^{N_2 - 1} \sum_{n_3 = 0}^{N_3 - 1}{e^{-i 2\pi \sum_{j = 1}^{3} \frac{k_j n_j}{N_j}} x[n_1, n_2, n_3]}
\end{equation}

where $n$ denotes the time index, $i$ is the imaginary unit, $k_d = 0,1, \ldots, N_d - 1$, and $X[k_1, k_2, \ldots, k_d]$ represents the frequency domain information at frequencies of $2 \pi k_j/N$ in dimension $j$ respectively.

The multidimensional Inverse Discrete Fourier Transform (IDFT) can also be given: 

\begin{equation}
\resizebox{0.9\columnwidth}{!}{$
x[n_1,n_2,n_3] = \frac{\sum_{k_1=0}^{N_1-1} \sum_{k_2=0}^{N_2-1} \sum_{k_3=0}^{N_3-1} e^{i 2\pi \sum_{j=1}^{3} \frac{n_j k_j}{N_j}}X[k_1,k_2,k_3]} {\prod_{l=1}^{3} N_l}
$}
\end{equation}

\subsection{Fourier Block for Hierarchical Segmentation Network}

\begin{equation}
\label{eq:block}
\begin{aligned}
    & \hat{X}^{l-1}_{spatial} = Padding(X^{l-1}_{spatial}) \\
    & W^l_{freq} \in \mathbb{C}^{C \times H \times W \times D}\ (Learnable\ Weight\ Matrix)\\
    & X^l_{freq} = DFT(LN(\hat{X}^{l-1}_{spatial}))\\
    & \hat{X}^l_{freq} = Crop(X^l_{freq} \odot W^l_{freq}) + Bias_{freq}\\
    & \hat{X}^{l}_{spatial} = MLP(LN(IDFT(\hat{X}^l_{freq}))) + X^{l-1}_{spatial}\\
    & X^{l}_{spatial} = Pooling(\hat{X}^{l}_{spatial})
\end{aligned}
\end{equation}

The given steps(Equation \ref{eq:block}) convert features between spatial and frequency domains using the Discrete Fourier Transform (DFT) and its inverse (IDFT). Here's a brief description of each step:

\subsubsection{Padding the Frequency Representation}
\begin{figure*}[]
  \centering
  \includegraphics[width=0.8\linewidth]{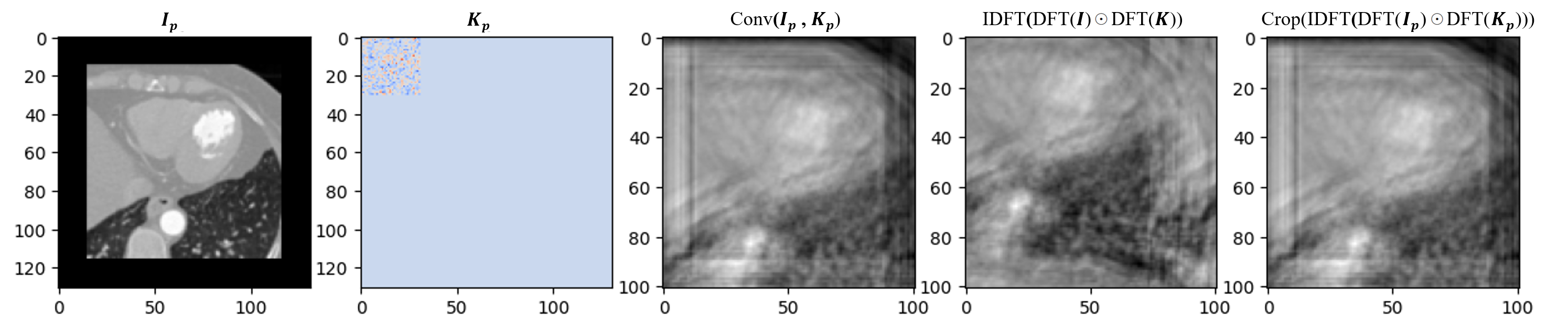}
  \caption{Convolution vs DFT-IDFT. $\textbf{I}_p$(or $\textbf{K}_p$) means image(or kernel) with padding, and $\odot$ means Hadamard product. The third figure denotes vanilla linear convolution result, while the latter two figures denote multiply in the frequency domain with(or without) padding respectively. A significant visual shift can be found when comparing the 3rd column with the 4th column.}
  \label{fig:conv}
\end{figure*}

The frequency representation($X^{l-1}_{freq}$) is padded to prevent wrap-around\cite{hunt1971matrix, pelkowitz1981frequency} effects (as shown in Fig. \ref{fig:conv}) in a subsequent convolution operation. The intuition behind this is that zero-padding the sequences gives enough "space" for the sequences to convolve linearly without the tail of one sequence wrapping around and interfering with the start of the convolution. The disturbed image may have a significant visual shift(3rd vs. 4th column in Fig. \ref{fig:conv}), and padding to a certain length effectively resists this interference(3rd vs. 5th column in Fig. \ref{fig:conv}). 

\subsubsection{Conversion to Frequency Domain} 
This equation suggests that the spatial representation of the $(l-1)$th layer, $\hat{X}^{l-1}_{spatial}$ undergoes a layer normalization (LN) before being transformed to the frequency domain using the Discrete Fourier Transform(DFT). 

\subsubsection{Convolution in the Frequency Domain with Bias Addition}
Element-wise multiplication (or Hadamard product, represented by $\odot$) between the frequency representation and the frequency-domain learnable weights($W^l_{freq}$) takes place. The shape of $W^l_{freq}$ is equivalent to $\hat{X}^{l-1}_{spatial}$ This is equivalent to convolution in the spatial domain. After the operation, the result is cropped to remove the padded regions and frequency-domain learnable bias is added.

\subsubsection{Conversion Back to Spatial Domain with Residual Connection}
The frequency representation is transformed back to the spatial domain using the inverse DFT (IDFT). This spatial data undergoes layer normalization (LN) and then passes through a multi-layer perceptron (MLP). The output is then added to the spatial representation from the previous $layerX^{l-1}_{spatial}$, forming a residual or skip connection.

\subsubsection{Pooling Operation in Spatial Domain}
The spatial representation is then downsampled using a pooling operation.

These steps demonstrate a mix of traditional convolutional neural network operations and spectral domain processing. Transforming between spatial and frequency domains can leverage the strengths of both representations in a neural network.

\subsection{Fourier Fusion Decoder}

\begin{equation}
\label{eq:decoder}
\begin{aligned}
    & E^{l-1}_{freq} = Crop_{inner}(DFT(E^{l-1}_{spatial})) \\
    & D^{l}_{freq} = Crop_{outer}(DFT(D^{l}_{spatial}))\\
    & D^{l-1}_{freq} \leftarrow D^{l}_{freq} + E^{l-1}_{freq} \\
    & D^{l-1}_{spatial} = IDFT(D^{l-1}_{freq})
\end{aligned}
\end{equation}

As outlined in the provided Equation \ref{eq:decoder}. the features derived from an encoder are represented as $E$, while those from a decoder are denoted by $D$. The superscripts $l-1$ and $l$ designate the depth of layers, respectively. The subscript 'spatial' and 'freq' are utilized to differentiate between features in the spatial domain and those in the frequency domain after undergoing a Fourier transformation, respectively.

A crucial procedure $Crop_{inner}$ is applied to the spatial features of the shallow encoder layer($E^{l-1}_{spatial}$) , which essentially isolates the low-frequency signals, thereby removing the high-frequency peripheries. In contrast, the $Crop_{outer}$ is applied to the spatial features of the deeper decoder layer($D^{l}_{spatial}$) , which retains the high-frequency details while discarding the inner low-frequency content.

Following this, a fusion operation is performed wherein the low-frequency semantic information from the encoder's shallow layer is combined with the high-frequency detail information from the decoder's deeper layer. The resultant spatial domain features($D^{l-1}_{spatial}$)  transformed by inverse discrete fourier transform(IDFT) serve as the input for the subsequent layer of the decoder. 

\begin{figure}[h]
  \centering
  \includegraphics[width=\linewidth]{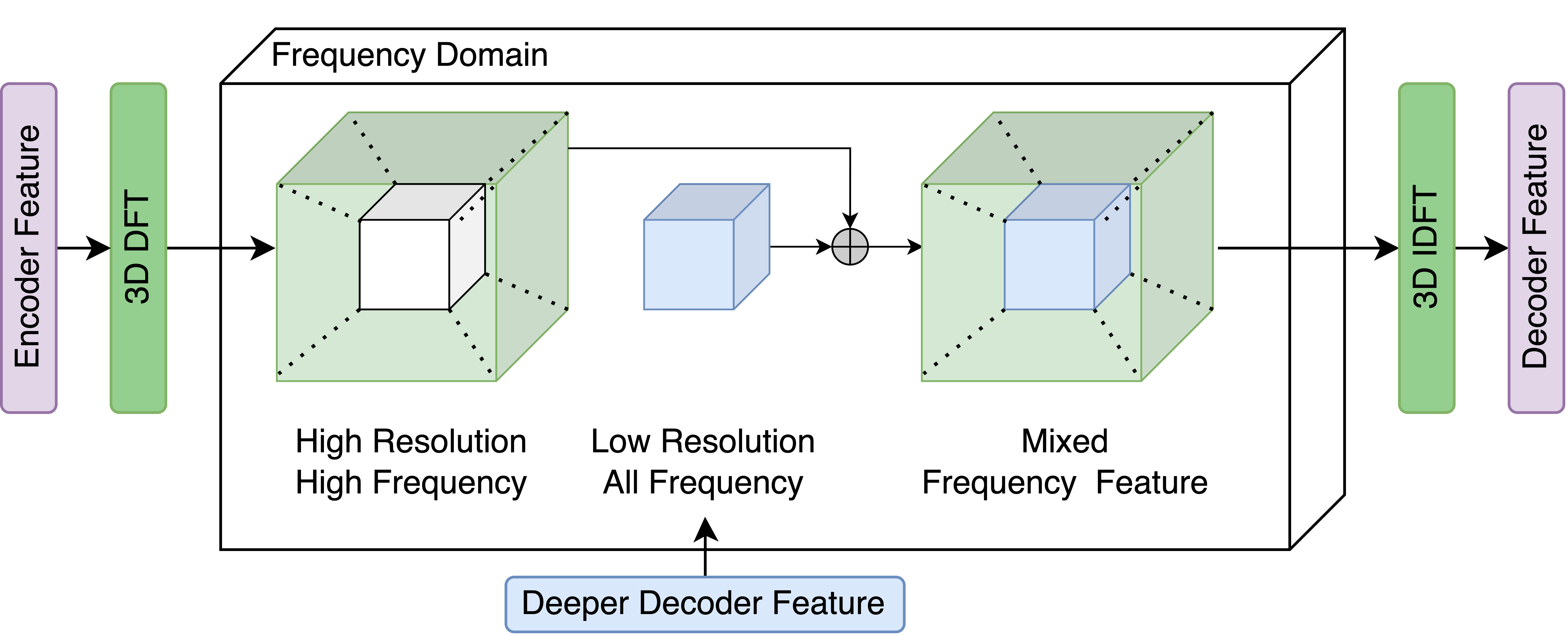}
  \caption{Fourier Fusion Decoder}
  \label{fig:decoder}
\end{figure}

The overall process of the Fourier fusion in the decoder is shown in Fig. \ref{fig:decoder}. In the shifted frequency spectrum, the central value represents low-frequency semantic information, while the edge value represents high-frequency detail information. This module preserves the high-frequency part of the frequency-domain features of the high-resolution features (detail information) and the frequency-domain features of the low-resolution features (semantic information) and then combines them. This approach is an advancement over the traditional method of direct concatenation, leveraging the frequency domain information to enhance the segmentation performance of the decoder.

\subsection{Loss Function}
We adopted a weighted combination of Dice loss and Cross-entropy loss as the loss function, which can be calculated by the following formula:

\begin{equation}
Loss = 1 - Dice - \frac{\lambda}{N} \sum_{i = 1}^{N} \sum_{k = 1}^{M} g_{ik} \log(p_{ik})
\end{equation}

where $N$ is the total number of voxels, $M$ is the total number of classes, $Dice$ is a metric whose calculation method is shown in the first formula in Section \ref{sec:Metrics}. $\lambda$ is the weight of cross-entropy loss, which we set 0.5 here. $g_{ik}$ takes 1 when the ground truth of voxel $i$ is class $k$ and 0 otherwise, and $p_{ik}$ is the probability that voxel $i$ is predicted to be class $k$.

\section{Experiments} \label{sec:experiment}

\subsection{Datasets}
Two cardiac tubular vessel segmentation datasets containing only coronary artery labels are used. One is a public dataset and one is an in-house dataset. The public ImageCAS dataset has 1000 3D CTA images from Guangdong Provincial People's Hospital, including only adult patients over 18 with a history of ischemic stroke, transient ischemic attack, and/or peripheral arterial disease. The Automatic Segmentation of Aorta and Coronary Arteries (ASACA) dataset contains two sub-datasets for evaluating vessel segmentation performance. ASACA100 and ASACA500 are the two sub-datasets, with the main difference being the number of CT images.
\subsection{Experienment Settings}
In this experiment, all images were inferenced with a sliding window (96, 96, 96), and a batch size of 4 was used for training. All metrics were evaluated without considering the background. The AdamW optimizer was employed with a weight decay value of $1e-6$. The learning rate is constant $1e-4$. The training, validation and test datasets were divided in ratio $8:1:1$. Our model was implemented in PyTorch and accelerated by 8 NVIDIA-A100 40GB GPUs. 

\begin{table}[]
\centering
\caption{Different Configuration of Fseg Network}
\label{tab:conf}
\begin{tabular}{ccc}
\hline
       & Feature Dimensions     & Number of Blocks \\ \hline
Fseg-S\textcolor{green}{\crossmark} & {[}12, 24, 48, 96{]}   & {[}2, 2, 4, 2{]} \\
Fseg-M\textcolor{orange}{\crossmark} & {[}24, 48, 96, 128{]}  & {[}2, 2, 4, 2{]} \\
Fseg-L\textcolor{red}{\crossmark} & {[}48, 96, 128, 256{]} & {[}2, 2, 4, 2{]} \\ \hline
\end{tabular}
\end{table}

Table \ref{tab:conf} presents the different configurations of the Fseg Network. The column labeled "Feature Dimensions" indicates the number of channels in the network across four stages(four grey downarrows in Fig \ref{fig:pipeline}). The values within the brackets represent the channel counts for each of the four stages in sequence. On the other hand, the "Number of Blocks" column specifies the number of encoder blocks utilized in each of the respective stages. For instance, the Fseg-S configuration, denoted by a green crossmark, has feature dimensions of [12, 24, 48, 96] and employs 2, 2, 4, and 2 encoder blocks in its four stages, respectively. Similarly, the configurations for Fseg-M (indicated by an orange crossmark) and Fseg-L (represented by a red crossmark) are also detailed.

\subsection{Data Augmentation}
The following data augmentation methods were used in this experiment to increase the variability of the training data. First, the intensity values of the input images were clipped to a specified range (-200 to 1000) and then mapped to a new range (0 to 1). Secondly, with a probability of 0.5, the input images and labels were randomly flipped along each spatial axis (x, y or z axis) and rotated by multiples of 90 degrees (up to 3 times). Thirdly, with a probability of 0.1, the intensity values of the input images were randomly scaled by a factor between 0.9 and 1.1, and with the same probability, they were randomly shifted by an offset between -0.1 and +0.1. These methods were used to increase the variability of the training data.

\subsection{Metrics} \label{sec:Metrics}
Dice score (Dice) and Intersection of Union (IoU) are used in our experiments to evaluate the accuracy of 3D segmentation. In the following section, let $X_{k}$ and $X_{k}^*$ be the set of volume pixels that are labeled as category $k$ of ground truth and prediction, where $|X_k \cap X_k^*|$ represents the counts of correctly predicted volume pixels. To calculate the surface distance, we defined $\partial X_k$ and $\partial X_k^*$ as the set of surfaces of ground truth and prediction perspectively where $x$ and $x^*$ denote a point of ground truth and prediction. Under this definition, Dice and IoU for each category $k$ be defined as:
\begin{equation}
    Dice = \frac{2|X_k \cap X_k^*|}{|X_k| + |X_k^*|}
\end{equation}

\begin{equation}
    IoU = \frac{|X_k \cap X_k^*|}{|X_k \cup X_k^*|}
\end{equation}

\subsection{Comparison}

\subsubsection{Compared to other recent State-of-the-Art approaches}

Analyzing the provided Table \ref{tab:sota}, which compares different neural network architectures across coronary datasets, several observations and conclusions can be drawn:

The proposed model, Fseg-L, consistently outperforms other models across all three datasets in terms of both Intersection over Union (IoU) and Dice coefficient. This suggests that Fseg-L is a robust model for the tasks at hand.

\begin{figure*}[h]
 \centering
 \includegraphics[width=0.8\linewidth]{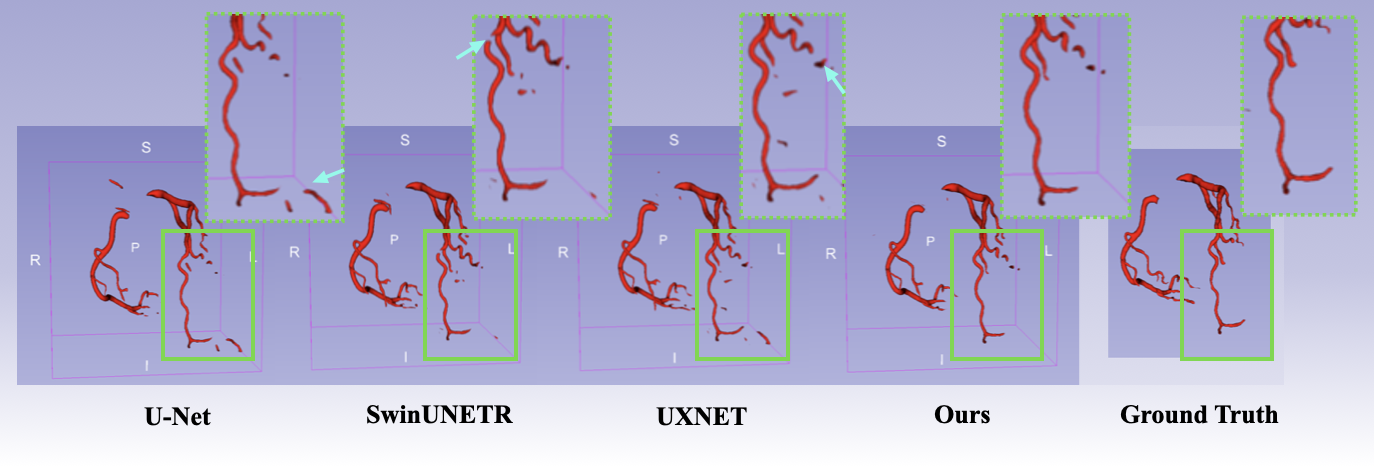}
 \caption{Comparison of segmentation results using some recent methods. A dashed box is an enlargement of a solid box for better comparison. The blue arrows indicate areas of poor segmentation.}
 \label{fig:comparision}
\end{figure*}

The 3D UXNet and Transunet models also show competitive performance across datasets. However, they still fall short when compared to Fseg-L. Its superior performance, combined with its stability (as indicated by the standard deviations), suggests that it is a promising model for segmentation tasks in the datasets considered.
Besides, some visual examples of segmentation results obtained by our model and the compared methods are shown in Fig. \ref{fig:comparision}, from which we can find our proposed method can obtain more accurate segmentation results.

\subsubsection{Computing Efficiency}

Table \ref{tab:flops} presents a comparison of various recent SOTA approaches on the ASACA500 dataset in terms of FLOPs (Floating Point Operations Per Second), the number of parameters, and the Dice coefficient. From the table, it's evident that the Fseg series, comprising of Fseg-S, Fseg-M, and Fseg-L, demonstrates competitive performance across all metrics. Notably:

\textbf{Efficiency}:
Fseg-S, with only 40.58G FLOPs, achieves a Dice coefficient of 0.8223. This is remarkable given that it outperforms U-Net, which requires over three times the computational cost (135.98G FLOPs) for a slightly lower Dice score of 0.8015.

\textbf{Scalability}:
As we scale from Fseg-S to Fseg-L, there's a consistent improvement in the Dice coefficient. Fseg-L, despite its high computational cost of 574.65G FLOPs, achieves the highest Dice score in the table at 0.8437. This suggests that the Fseg architecture scales well with increased complexity.

\begin{table}[]
\centering
\caption{Different Approaches Comparision on Coranary Segmentation Task}
\label{tab:sota}
\begin{tabular}{@{}ccll@{}}
\toprule
Dataset                   & Network       & \multicolumn{1}{c}{IoU} & \multicolumn{1}{c}{Dice} \\ \midrule
\multirow{7}{*}{ASACA100} & Unet\cite{cciccek20163d}          & $0.6563\pm0.0142$       & $0.7963\pm0.0068$        \\
                          & Unetr\cite{hatamizadeh2022unetr}         & $0.6832\pm0.0242$       & $0.8132\pm0.0049$        \\
                          & Swinunetr\cite{hatamizadeh2021swin}     & $0.6932\pm0.0106$       & $0.8232\pm0.0133$        \\
                          & nnFormer\cite{zhou2021nnformer}      & $0.6791\pm0.0216$       & $0.8091\pm0.0156$        \\
                          & Transunet\cite{chen2021transunet}     & $0.6923\pm0.0079$       & $0.8123\pm0.0234$        \\
                          & 3D UXNet\cite{lee20223d}         & $0.6952\pm0.0160$       & $0.8252\pm0.0051$        \\
                          & \textbf{Fseg-L}(\textbf{Ours}) & $\textbf{0.7123}\pm0.0245$       & $\textbf{0.8323}\pm0.0048$        \\ \midrule
\multirow{7}{*}{ASACA500} & Unet\cite{cciccek20163d}          & $0.6712\pm0.0244$       & $0.8015\pm0.0052$        \\
                          & Unetr\cite{hatamizadeh2022unetr}         & $0.6873\pm0.0089$       & $0.8132\pm0.0220$        \\
                          & SwinUNETR\cite{hatamizadeh2021swin}     & $0.7054\pm0.0195$       & $0.8260\pm0.0155$        \\
                          & nnFormer\cite{zhou2021nnformer}      & $0.6612\pm0.0076$       & $0.8061\pm0.0209$        \\
                          & Transunet\cite{chen2021transunet}     & $0.7103\pm0.0143$       & $0.8220\pm0.0153$        \\
                          & 3D UXNet\cite{lee20223d}         & $0.7065\pm0.0133$       & $0.8365\pm0.0158$        \\
                          & \textbf{Fseg-L}(\textbf{Ours}) & $\textbf{0.7137}\pm0.0152$       & $\textbf{0.8437}\pm0.0070$        \\ \midrule
\multirow{7}{*}{ImageCAS} & Unet\cite{cciccek20163d}          & $0.6588\pm0.0071$       & $0.7788\pm0.0155$        \\
                          & Unetr\cite{hatamizadeh2022unetr}         & $0.6604\pm0.0058$       & $0.7804\pm0.0153$        \\
                          & SwinUNETR\cite{hatamizadeh2021swin}     & $0.6716\pm0.0107$       & $0.7916\pm0.0040$        \\
                          & nnFormer\cite{zhou2021nnformer}      & $0.6442\pm0.0144$       & $0.7842\pm0.0171$        \\
                          & Transunet\cite{chen2021transunet}     & $0.6502\pm0.0215$       & $0.7902\pm0.0071$        \\
                          & 3D UXNet\cite{lee20223d}         & $0.6566\pm0.0053$       & $0.7966\pm0.0224$        \\
                          & \textbf{Fseg-L}(\textbf{Ours}) & $\textbf{0.6732}\pm0.0152$       & $\textbf{0.8032}\pm0.0113$        \\ \midrule
\end{tabular}
\end{table}

\begin{table}[]
\centering
\caption{Ablation study with fusion and global filter mechanism}
\label{tab:ablation}
\begin{tabular}{@{}cccc@{}}
\toprule
Decoder                 & Filter            & Padding & Dice              \\ \midrule
\multirow{3}{*}{Skip}   & dwconv 7*7                     & None    & $0.8365\pm0.0158$ \\ \cmidrule(l){2-4} 
                        & \multirow{2}{*}{Fourier} & None    & $0.8405\pm0.0105$ \\
                        &                          & Padding & $0.8427\pm0.0168$ \\ \midrule
\multirow{3}{*}{Fusion} & dwconv 7*7                     & None    & $0.8401\pm0.0135$ \\ \cmidrule(l){2-4} 
                        & \multirow{2}{*}{Fourier} & None & $0.8435\pm0.0119$ \\
                        &                          & Padding & $\textbf{0.8437}\pm0.0061$ \\ \bottomrule
\end{tabular}
\end{table}

\begin{table}[]
\centering
\caption{Comparison of recent SOTA approaches on the ASACA500 dataset}
\label{tab:flops}
\begin{tabular}{@{}cccl@{}}
\toprule
          & FLOPs   & \#Params & Dice \\ \midrule
U-Net\cite{cciccek20163d}     & 135.98G & 4.81M    & 0.8015                   \\
UNETR\cite{hatamizadeh2022unetr}     & 82.6G   & 92.8M    & 0.8132                   \\
SwinUNETR\cite{hatamizadeh2021swin} & 328.41G & 62.23M   & 0.8260                   \\
nnFormer\cite{zhou2021nnformer}  & 240.26G & 149.34M  & 0.8061                   \\
3D UX-Net\cite{lee20223d} & 639.45G & 53.02M   & 0.8365                   \\
Fseg-S\textcolor{green}{\crossmark}    & 40.58G  & 27.14M   & 0.8223                   \\
Fseg-M\textcolor{orange}{\crossmark}    & 148.17G & 41.60M   & 0.8283                   \\
Fseg-L\textcolor{red}{\crossmark}    & 574.65G & 80.24M   & \textbf{0.8437}                   \\ \bottomrule
\end{tabular}
\end{table}

\subsubsection{Ablation study for decoder fusion and global filter}

The table \ref{tab:ablation} presents an ablation study examining the effects of different decoders, filters, and padding mechanisms on the Dice coefficient, with "Skip" and "Fusion" decoders being evaluated alongside two filter types: "dwconv 7*7 (depthwise convolution with kernel size 7)" and "Fourier". For each filter type, results are shown both with and without padding. From the data, it's evident that the Fusion decoder consistently outperforms the Skip decoder when paired with the same filter and padding settings. Moreover, the Fourier filter, especially when combined with padding, tends to achieve higher Dice coefficients compared to the dwconv 7*7 filter. This observation is further reinforced by the fact that padding has a positive effect on the Dice coefficient when the Fourier filter is employed, leading to a slight improvement in scores for both decoders.  Based on the presented data, the Fusion decoder combined with the Fourier filter and padding appears to be the most effective configuration for maximizing the Dice coefficient.

The analysis demonstrates the trade-offs between computational complexity and segmentation performance in various architecture configurations. The results suggest that deeper networks with larger channels generally lead to better segmentation accuracy, but incorporating FFT-based techniques can offer competitive results with reduced computational demands.

\section{Conclusion}

In this research, we have introduced a novel approach to 3D vessel segmentation by harnessing the power of frequency domain learning. This method not only ensures the preservation of global interactions and anti-aliasing properties within the network but also addresses computational constraints, making it a viable alternative to traditional attention mechanisms. Our unique zero-parameter decoder, constructed through the fusion of different frequency components, maximizes the modeling capability, offering a more efficient means of integrating features between the encoder and decoder. Experimental results on both public and in-house datasets underscore the superiority of our approach, as it outperforms other recent methods in 3D vessel segmentation tasks. By leveraging the efficiency of the Fast Fourier Transform, we have successfully reduced the computational demands of the network without compromising its global receptive field. This study paves the way for more efficient and accurate computer-aided diagnostic systems, especially in the realm of coronary microvascular disease detection and intervention.

\bibliography{ref}

\end{document}